\documentclass[twocolumn,aps,prl,groupedaddress,showpacs]{revtex4}
\usepackage{epsfig,amssymb,amsmath}

\def\comment#1{}\def\labell#1{\label{#1}}
\begin{document}
\title{Robust strategies for lossy quantum interferometry}
\author{Lorenzo Maccone and Giovanni De Cillis}\affiliation{ QUIT -
  Quantum Information Theory Group, Dipartimento di Fisica ``A.
  Volta'' Universit\`a di Pavia, via A.  Bassi 6, I-27100 Pavia,
  Italy.}

\begin{abstract}
  We give a simple multiround strategy that permits to beat the shot
  noise limit when performing interferometric measurements even in the
  presence of loss. In terms of the average photon number employed,
  our procedure can achieve twice the sensitivity of conventional
  interferometric ones in the noiseless case. In addition, it is more
  precise than the (recently proposed) optimal two-mode strategy even
  in the presence of loss.
\end{abstract}
\maketitle 
The shot noise limit is the minimum noise level that the Heisenberg
uncertainty relations permit to achieve when classical states are
employed in the apparatuses. Many quantum strategies have been
proposed to beat the shot noise~\cite{review,qmetr} and to achieve the
ultimate Heisenberg limit~\cite{interf}, but virtually all of them are
very sensitive to noise and loss of photons~\cite{gerry}.  Only very
recently some interferometric strategies were presented that can beat
the shot noise even in the presence of relevant losses of
photons~\cite{walmsley,dowling}. These are all instances of parallel
strategies~\cite{qmetr}\comment{This is not so rigorous: parallel
  strategies are those where all the probes are evolved by the
  generator of the transformations: here only the upper arm is
  evolved, not the reference arm}, where both arms of the
interferometer are sampled at the same time using a mode-entangled
quantum state of the light (see Fig.~\ref{f:strategy}a). In addition
to the parallel strategies, in quantum metrology it is also possible
to achieve the Heisenberg limit using multiround (or sequential)
strategies~\cite{luis,nature} which, in the noiseless case, are
equivalent in terms of resources and of achievable
precision~\cite{qmetr}.

\begin{figure}[hbt]
\begin{center}
\epsfxsize=1.\hsize\leavevmode\epsffile{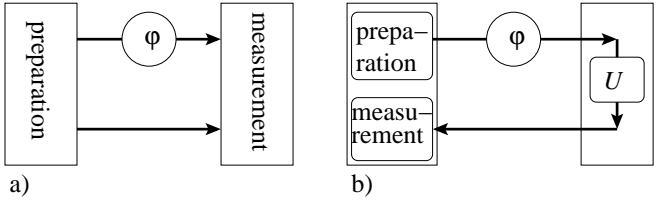}
\end{center}
\vspace{-.5cm}
\caption{a) Parallel strategies for interferometry. Both arms of the
  interferometer are sampled at the same time using a two-mode
  entangled state. The phase factor $\varphi$ is imprinted in the
  state as a phase difference between the two modes. b) Sequential
  strategy proposed here. A loss-resistant single mode state is sent
  through the first interferometer arm, sampling the phase $\varphi$.
  Then a unitary transformation is applied and the state is sent {\em
    back} through the other interferometer arm. This is needed so that
  the final phase shift experienced by the state is only the
  relative phase in the interferometer. The state is measured after
  one (or more) round trips.}
\labell{f:strategy}\end{figure}

Here we detail how multiround strategies can be used to perform
interferometry--- see Fig.~\ref{f:strategy}b. An appropriate input
state is prepared (we will analyze two examples below). This state is
fed into the first interferometer arm. It picks up a phase
$\varphi+\vartheta$, where $\varphi$ is the interferometric phase we
want to estimate, and $\vartheta$ is the absolute phase picked up by
the free evolution in the arm (which is equal to the phase which would
be picked up also in the reference arm). The main trick of multiround
interferometry is the use of the unitary
\begin{eqnarray}
U=\sum_{n=0}^M|M-n\rangle\langle
n|+\sum_{n=M+1}^{\infty}|n\rangle\langle n|
\labell{defu}\;,
\end{eqnarray}
where $|n\rangle$ is the Fock basis and $M$ is the largest nonzero
component of the initial input state. The purpose of this unitary is
to permute the first $M$ components of the Fock state expansion of a
state in such a way that, when the state is sent back through the
reference arm, the absolute phase $\vartheta$ is removed from the
state (only an irrelevant global phase factor $e^{iM\vartheta}$
persists). Thus, at the end of the round trip of
Fig.~\ref{f:strategy}b (multiple round trips are also possible), only
the relative phase $\varphi$ is imprinted on the state. A measurement
is finally performed to estimate this phase.

This multiround interferometry employs the same average energy and the
same modes as the conventional (parallel) strategies, but it can
achieve twice the sensitivity and it is more robust against noise.  In
fact, we show that, with an appropriate choice of inputs, our protocol
permits to estimate the phase with an error which is smaller than what
is achieved by the strategy detailed in Ref.~\cite{walmsley}, which is
claimed to be the optimal two-mode strategy. In the presence of loss,
this is not unexpected, as parallel protocols rely on entanglement
which is notoriously fragile to noise. Instead, in the noiseless case,
this can be seen easily with a simple example.  Recall that the
Heisenberg limit~\cite{interf} is essentially an application of the
time-energy uncertainty, $\Delta\varphi\Delta h\geqslant 1/2$, where
$h$ is the generator of the unitary that inserts the phase $\varphi$
into the system~\cite{caves}, namely $h=a^\dag a$ ($a$ being the
annihilation operator of the first arm of the interferometer). Optimal
two-mode states, such as the NOON state
$(|N0\rangle+|0N\rangle)/\sqrt{2}$, have $\Delta h=N/2$. However, the
corresponding single-mode ``NO'' state of same average number of
photons, namely the state $(|2N\rangle+|0\rangle)/\sqrt{2}$, has
$\Delta h=N$. Both states achieve the Heisenberg-limited sensitivity
of $\Delta\varphi=1/(2\Delta h)$, but in terms of the average number
of photons $N$, the NOON state can achieve $\Delta\varphi_{NOON}=1/N$,
whereas the NO state can achieve $\Delta\varphi_{NO}=1/(2N)$,
i.e.~twice the sensitivity. The NO state is, however, just as
sensitive to noise as the NOON state: the loss of a single photon
renders both states useless to phase estimation.

The rest of the paper is devoted to presenting two examples of
loss-resistant multiround interferometry based on two different input
states, the optimal phase state and the single-mode M\&M state. Both
are able to beat the optimal parallel strategy for some values of
parameters.  

We start by considering the single-mode optimal phase
state $|\psi_{opt}\rangle$ introduced in~\cite{wisemanph,buzek} (and
later extended to the multi-mode case in~\cite{berrywise}, building
on~\cite{barry}), i.e.~the state
\begin{eqnarray}
  |\psi_{opt}\rangle\equiv \sqrt{\frac
    2{M+1}}\sum_{n=0}^M\sin\Big(\frac{\pi(n+1/2)}{M+1}\Big)
|n\rangle
\labell{optimaldef}\;,
\end{eqnarray}
where $M$ is a parameter identifying the average number of photons
$N=M/2$. In the noiseless case, this state can be used to achieve the
ultimate precision~\cite{interf} in the absolute phase estimation,
i.e.~the Heisenberg scaling $\sim 1/N$. Moreover, this state is highly
robust to loss, since the loss tends to deplete primarily the
components with large Fock numbers $n$ which are not very populated in
this state. Following the scheme of Fig.~\ref{f:strategy}b, this state
is sent through the first arm of the interferometer, where it is
subject to a phase shift of $\varphi+\vartheta$, where $\varphi$ is
the interferometric phase we want to estimate, and $\vartheta$ is the
absolute phase picked up by the free evolution in the arm. During this
transit, the state is also evolved by the loss map, described by the
Kraus operators $K_i\equiv(\eta^{-1}-1)^{i/2}a^i\eta^{a^\dag
  a/2}/\sqrt{i!}$, where $a$ is the annihilation operator of the
optical mode and $\eta$ is its transmissivity or quantum efficiency.
(Note that the loss and the phase accumulation commute, so that the
order in which we apply these two transformations is irrelevant.)
Then, the state is subject to the unitary evolution $U$ of
Eq.~(\ref{defu}) and it is sent back along the reference arm.  Thanks
to the permutation of the Fock components that $U$ applies to the
state, the free evolution is effectively reversed, so that the
absolute phase $\vartheta$ that was picked up in the first arm is
removed while the radiation travels back through the reference arm.
Also in the reference arm the state is typically subject to the loss
(although there are interesting cases where the reference may be
considered noiseless). Finally, the state is subject to the
measurement. For the optimal phase state, a good
measurement~\cite{buzek} can be obtained by considering the orthogonal
POVM composed by the projectors on the Pegg-Barnett states
\begin{eqnarray}
  |\Phi_l\rangle\equiv\frac 1{\sqrt{M+1}}
  \sum_{n=0}^Me^{i\Phi_l}|n\rangle\,
  \ \mbox{ with }\Phi_l=\frac{2\pi l}{M+1},
\labell{defpovm}\;
\end{eqnarray}
with $l=0,\cdots,M$. The RMS of the probability distribution obtained
from this POVM is a function of the interferometer phase $\varphi$.
Its minimum value gives the minimum error that our interferometer can
achieve, which is plotted as the continuous lines in
Fig.~\ref{f:optimal}. In addition, we have also directly estimated the
error through the Holevo variance~\cite{holevo}
$\Delta\Phi=({S_\Phi^{-2}-1})^{\frac 12}$, where $S_\Phi=|\langle
e^{i\Phi}\rangle|$ is the average value of the function $e^{i\Phi}$
weighted with the probability obtained from the continuous POVM
$|\Phi\rangle\langle\Phi|\;d\Phi$, obtained using the phase states of
Eq.~(\ref{defpovm}) for arbitrary $\Phi$. The Holevo variance is more
appropriate than the RMS for the estimation of the phase error, since
the phase is a periodic quantity~\cite{holevo}. However, as is clear
from our plots, the Holevo variance is well approximated by the RMS
when these two quantities are small enough (compared to $2\pi$).

It is tedious but straightforward to calculate that, after a round
trip characterized by the same quantum efficiency $\eta$ in both arms,
the state $|\psi_{opt}\rangle$ is evolved into
\begin{eqnarray}
  \rho=\frac 2{M+1}\sum_{i,j=0}^M(1-\eta)^{i+j}\eta^{M-j}
  \sum_{n,m}\omega_n\omega_me^{i\varphi(m-n)}|n\rangle\langle m|,&&
  \nonumber
  \\\nonumber
  \omega_k\equiv{\left(\begin{matrix}\scriptstyle
        k+j\cr\scriptstyle
        j\end{matrix}\right)^{\frac 12}
    \left(\begin{matrix}\scriptstyle M-k-j+i\cr\scriptstyle
        i\end{matrix}\right)}^{\frac 12}
\sin\Big[
\tfrac{\pi(M-k-j+i+\frac 12)}{M+1}\Big]&&
\end{eqnarray}
where the second sum runs between max$(0,i-j)$ and $M-j$.
\comment{Incidentally, note that the normalization condition of this
  state gives a nice formula (se c'e' spazio si puo' aggiungere la
  formula di MR8) } From this state, one can estimate the probability
distribution of the POVM of Eq.~(\ref{defpovm}), namely
\begin{eqnarray}
p(l)=\frac 2{(M+1)^2}\sum_{i,j}(1-\eta)^{i+j}\eta^{M-j}
\Big|\sum_n\omega_n\;e^{-in(\varphi+\Phi_l)}
\Big|^2.
\nonumber
\end{eqnarray}
 The error in the estimation of the phase from a
measurement on $\rho$ is given by the RMS of $p(l)$, plotted in
Fig.~\ref{f:optimal} as a function of the average photon
number $N$ and quantum efficiency $\eta$.

\begin{figure}[hbt]
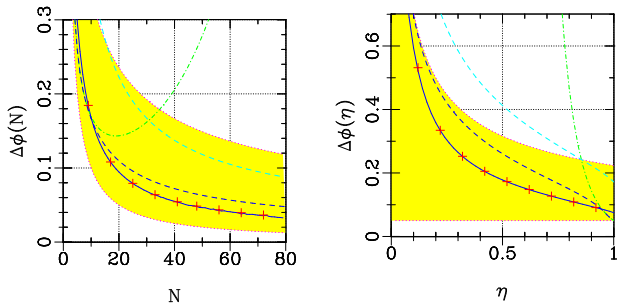

\begin{center}
\includegraphics[scale=0.3]{enne.eps}
\hspace{3mm}
\includegraphics[scale=0.3]{eta.eps}
\end{center}
\vspace{-.5cm}
\caption{Left) Solid line: error in the estimation of the phase from
  the state $\rho$, i.e.~minimum (over $\varphi$) of the RMS of the
  probability $p(l)$, as a function of the average number of photons
  $N$ for $\eta=0.9$. Lower dashed line: plot of the error from the
  optimal two-mode lossy interferometry from~\cite{walmsley}. Our
  single-mode method achieves a higher precision for a wide range of
  parameters.  Plus signs: Holevo variance $\Delta\Phi$, which closely
  tracks the minimum RMS.  Dot-dash line: behavior of the two mode
  state $(|N0\rangle+|0N\rangle/\sqrt{2}$, which is optimal only for
  high $\eta$ and low $N$. Upper grey dashed line: average over
  $\varphi$ of the RMS. The gray shaded area encloses the ``quantum''
  trajectories, i.e.~the ones included between the shot noise limit
  $1/\sqrt{N\eta}$ and the Heisenberg limit $1/N$.  Right) Same as the
  previous, but as a function of $\eta$ for $N=20$.}
\labell{f:optimal}\end{figure}

The second input state we consider is a single-mode analogous of the
M\&M state introduced in~\cite{dowling}, i.e. the state
\begin{eqnarray}
(|M\rangle+|M'\rangle)/\sqrt{2}\labell{mmdef}\;,
\end{eqnarray}
where $M>M'$
and whose average photon number is $N=(M+M')/2$. Again, it is
straightforward to obtain the output state, after the round trip of
Fig.~\ref{f:strategy}b:
\begin{eqnarray}
  \sigma&=&
  \textstyle
  \sum_{j=-\delta}^{M'}\alpha_j|j+\delta\rangle\langle j+\delta|+
  \sum_{j=0}^{M}\beta_j|j\rangle\langle j|
  \labell{rh}\\\nonumber
  &&+\textstyle
  \frac
  12\sum_{j=0}^{M'}\gamma_j\big[e^{-i\delta\varphi}|j+\delta\rangle\langle
  j|+e^{i\delta\varphi}|j\rangle\langle j+\delta|\big]
  \;,\end{eqnarray}
\begin{eqnarray}
  \alpha_j&=&\textstyle\sum_{i=max(0,j)}^{M'}f_{ij}
  \left(\begin{matrix} 
      \scriptstyle M'\cr\scriptstyle i\end{matrix}\right) \left(\begin{matrix}
      \scriptstyle  i+\delta\cr\scriptstyle i-j\end{matrix}\right)/2,
  \\\beta_j&=&\textstyle\sum_{i=j}^{M}f_{ij}
  \left(\begin{matrix} \scriptstyle M\cr\scriptstyle
      i\end{matrix}\right) \left(\begin{matrix}
      \scriptstyle i\cr\scriptstyle j\end{matrix}\right)/2, 
  \\\gamma_j&=&\textstyle\sum_{i=j}^{M'}f_{ij}
  \Big[\left(\begin{matrix} \scriptstyle M'\cr\scriptstyle
      i\end{matrix}\right)\left(\begin{matrix} \scriptstyle M\cr\scriptstyle
      i\end{matrix}\right)\left(\begin{matrix} \scriptstyle
      i+\delta\cr\scriptstyle
      i-j\end{matrix}\right) \left(\begin{matrix}
      \scriptstyle i\cr\scriptstyle j\end{matrix}\right)\Big]^{\frac 12},
\labell{ll}\;
\end{eqnarray}
with $\delta\equiv M-M'$ and
$f_{ij}\equiv(1-\eta)^{2i-j}\eta^{M-i+j}$. To extract the phase from
the state $\sigma$, in analogy to the two-mode case of~\cite{dowling},
we can measure the observable 
\begin{eqnarray}
  A=\sum_{k=0}^{M'}|M-k\rangle\langle
  M'-k|+|M'-k\rangle\langle M-k|\labell{defa}\;.
\end{eqnarray}
Then, the error on the phase can be obtained from the RMS of $A$ using
error propagation, namely
\begin{eqnarray}
  \Delta\varphi=\Delta A/\big|\frac\partial{\partial\varphi}\langle
  A\rangle\big|=\frac{\sqrt{\Theta-\cos^2(\delta\varphi)\Gamma^2}}
  {\delta|\sin(\delta\varphi)\Gamma|}
\labell{err}\;,
\end{eqnarray}
where $\Theta=\sum_{k=0}^{M'}\alpha_k+
\alpha_{k-\delta}+\beta_k+\beta_{k+\delta}$ and
$\Gamma=\sum_{k=0}^{M'}\gamma_k$. This quantity is plotted in
Fig.~\ref{f:mm} from which it is evident that, also in this case, the
multiround protocol can achieve a better sensitivity than the optimal
two-mode one. The single-mode M\&M state can take advantage of most of
the implementation ideas presented in~\cite{dowling} for the two-mode
case. It may appear surprising that the state $\sigma$ permits to
achieve a greater precision than the NOON state
$(|N0\rangle+|0N\rangle)/\sqrt{2}$ even for high values of $\eta$. As
discussed above, this is essentially due to the fact that a single
mode state performs better than a two-mode state in terms of the
average number of photons $N$.

\begin{figure}[hbt]
\begin{center}
\includegraphics[scale=0.3]{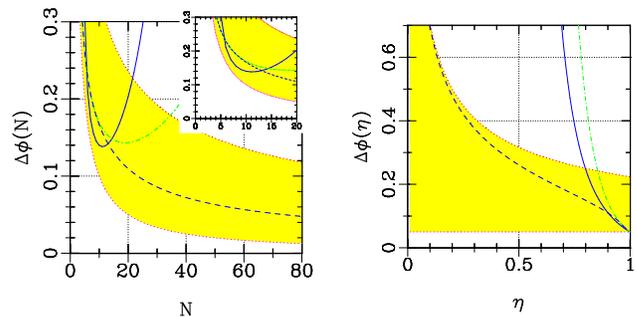}
\hspace{3mm}
\includegraphics[scale=0.3]{mmeta.eps}
\end{center}
\vspace{-.5cm}
\caption{Left) Solid line: error in the estimation of the phase from
  the state $\sigma$ of Eq.~(\ref{rh}), i.e.~minimum (over $\varphi$)
  of the function $\Delta\varphi$ of Eq.~(\ref{err}), as a function of
  the average number of photons $N$ for $\eta=0.9$. Here $M'=3$ and
  $M=2N-3$. Dashed line: plot of the error from the optimal two-mode
  lossy interferometry from~\cite{walmsley}.  Again, as in
  Fig.~\ref{f:optimal}, our single-mode method achieves a better
  sensitivity in some range. Dot-dash line: behavior of the state
  $(|N0\rangle+|0N\rangle/\sqrt{2}$. The inset is an enlargement for
  small values of $N$. Right) Same as the previous, but as a function
  of $\eta$ for $N=20$, $M=30$, $M'=10$.}
\labell{f:mm}\end{figure}

In conclusion, we have given a strategy for determining the relative
phase in an interferometer using single-mode states that are sent
through the interferometer in a round trip, interleaved by the unitary
$U$ of Eq.~(\ref{defu}). This entails that a) in the noiseless case a
double sensitivity can be reached over the optimal two-mode states
(such as the NOON state) in terms of the average number of photons, b)
in the lossy case we can achieve a better phase sensitivity than what
is claimed to be the optimal two-mode strategy~\cite{walmsley},
proving that multiround protocols are preferable in the presence of
noise. The robustness in the face of loss stems from two main
properties. On one side there is no entanglement between different
modes, and it is well known that entanglement is very sensitive to
noise~\cite{gerry}. On the other side, the fact that we are using a
single mode permits to double the phase sensitivity over the two-mode
entangled case, since all the photons travel through one mode only.
One may object that the increased phase sensitivity arises because we
are devoting more resources to the estimation. This objection is
unfounded since the average number of photons employed in the two
strategies is the same, $N$. One cannot even say that in the two-mode
strategies the phase $\varphi$ is sampled by less photons, because the
number of photons that travel through an arm of an interferometer is
an undefined quantity (the ``which path'' information is complementary
to the phase information). One can only bound the number of photons
traveling through each arm with the total number of photons, $N$,
injected in the interferometer.

\acknowledgments We thank R. Demkowicz-Dobrzanski for having kindly
provided the data of the optimal two-mode state of~\cite{walmsley}.


\begin{references}
\bibitem{review} For a recent review, see V. Giovannetti, S. Lloyd,
  and L. Maccone, Science {\bf 306}, 1330 (2004).
\bibitem{qmetr} V. Giovannetti, S. Lloyd, L. Maccone, Phys. Rev. Lett.
  {\bf 96}, 010401 (2006); S. L. Braunstein, Nature {\bf 440}, 617
  (2006).
\bibitem{gerry}G. Gilbert, M. Hamrick, Y.S. Weinstein, J.Opt. Soc. Am.
  B, {\bf 25,} 1336 (2008).
\bibitem{walmsley}U. Dorner, R. Demkowicz-Dobrzanski, B. J. Smith, J.
  S. Lundeen, W. Wasilewski, K. Banaszek, I. A. Walmsley,
  arXiv:0807.3659 [quant-ph] (2008).
\bibitem{dowling}S.D. Huver, C.F. Wildfeuer, J.P. Dowling,
  arXiv:0805.0296 [quant-ph] (2008).
\bibitem{luis}A. Luis, Phys. Rev. A {\bf 65}, 025802 (2002).
\bibitem{nature}B.L. Higgins, D.W. Berry, S.D. Bartlett, H.M.
  Wiseman, and G.J. Pryde, Nature {\bf 450}, 393 (2007).
\bibitem{wisemanph}H.M. Wiseman and R.B. Killip, Phys. Rev. A {\bf
    56,} 944 (1997).
\bibitem{buzek}V. Bu\v zek, R. Derka, and S. Massar, Phys. Rev.
  Lett. {\bf 82,} 2207 (1999).
\bibitem{berrywise}D.W. Berry and H.M. Wiseman, Phys. Rev. Lett. {\bf
    85,} 5098 (2000).
\bibitem{barry}B.C. Sanders and G.J. Milburn, Phys. Rev. Lett. {\bf
    75,} 2944 (1995); B.C. Sanders, G.J. Milburn, and Z. Zhang, J.
  Mod. Opt. {\bf 44}, 1309 (1997).
\bibitem{interf} J. J. Bollinger, W. M. Itano, D. J.  Wineland, and D.
  J. Heinzen, Phys. Rev. A {\bf 54}, R4649 (1996).
\bibitem{holevo}A.S. Holevo, {\em Probabilistic and statistical
    aspects of quantum theory} (North Holland pub. co., Amsterdam,
  1982).
\bibitem{caves}S. L. Braunstein, C. M. Caves and G. J. Milburn, Ann.
  Phys. {\bf 247}, 135 (1996); S. L. Braunstein, C. M. Caves, Phys.
  Rev. Lett. {\bf 72}, 3439 (1994).
\end{references}
\end{document}